\documentclass[amssymb,prl,twocolumn,showpacs]{revtex4}
\usepackage{amsmath}
\usepackage{bbm}
\usepackage{graphicx}

\begin{document}
\title{\bf 
Resonance Fluorescence in Transport through Quantum Dots: Noise Properties}
\author{Rafael S\'anchez$^1$, Gloria Platero$^1$ and Tobias Brandes$^2$}
\affiliation{
1-Instituto de Ciencia de Materiales, CSIC, Cantoblanco, Madrid, 28049, Spain. \\
2-Institut f\"ur Theoretische Physik, Technische Universit\"at Berlin, D-10623 Berlin, Germany.
}
\date{\today}
\begin{abstract}

We study a two-level quantum dot embedded in a phonon bath and irradiated by a time-dependent ac field and develope a method that allows us to extract simultaneously the full counting statistics of the electronic tunneling and relaxation (by phononic emission) events  as well as their correlation. We find that the quantum noise of both the transmitted electrons and the emitted phonons can be controlled by the manipulation of external parameters such as the driving field intensity or the bias voltage. 

\end{abstract}
\maketitle

The complete knowledge of the statistics and, in concrete, the properties of the fluctuations of the number of particles emitted from a quantum system has been a topic of intense studies in Quantum Optics \cite{optics,djuricSearch}
and, in more recent years, in Quantum Transport\cite{transportfcs,transportnoise}. In particular, {\it purely quantum} features like an anti-bunching of photons emitted from a closed two-level atom under a resonant field\cite{mandel}, or a bunching of electrons tunneling through interacting two-levels quantum dots (QD)\cite{superPoisson} have been reported. 
The electronic case receives special interest since it has been recently adressed for many different physical systems such as single electron tunneling
devices\cite{eth,uts,kiesslich06,djuricJAP,groth,sukhorukov}, molecules\cite{thielmann}, charge shuttles\cite{pis}, surface
acoustic waves driven single electron pumps\cite{rob} or beam
splitting configurations\cite{che}.
Here, we show that the {\em combined}
statistics of Fermions and Bosons is a very sensitive tool for
extracting information from time-dependent, driven systems.
In particular, phonon emission has been measured by its influence on the electronic current in two-level systems\cite{ramon}. We analyze the electron and phonon noises and find that they can be tuned back and forth between sub- and super-Poissonian character by using the strength of an ac driving field or the bias voltage. 
For this purpose, we develop a general method 
to  simultaneously extract the full counting statistics of
single electron tunneling  and (phonon mediated) relaxation events. 

\begin{figure}[htb]
\begin{center}
\includegraphics[width=1.9in,clip]{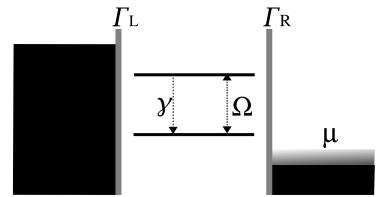}
\end{center}
\caption{\label{esquema}\small A two-level QD coupled to two electronic leads. Electrons entering the QD from the left oscillate between both levels with Rabi frequency $\Omega$, can relax with rate $\gamma$, or tunnel out if its energy is greater than $\mu$.
}
\end{figure}

Our system consists of a two level quantum dot (QD) connected to two fermionic leads by tunnel barriers cf. Fig. \ref{esquema}. 
The Coulomb repulsion inside the QD is assumed to be so large that only single occupation is allowed ({\it Coulomb blockade} regime). The lattice vibrations induce, at low temperatures, inelastic transitions from the upper to the lower state.
In analogy to Resonance Fluorescence (RF) in quantum optics,
a time-dependent ac field with a frequency $\omega$  drives the transition between the two levels $\varepsilon_1$, $\varepsilon_2$ close to resonance,
$\Delta_\omega=\varepsilon_2-\varepsilon_1-\omega\approx 0$, which allows us to assume the rotating wave approximation.
Thus, the electron in the QD is coherently delocalized between both levels performing {\it photon-assisted Rabi oscillations}\cite{gloria}. For simplicity, we consider spinless electrons.

The components of the density matrix $\rho (t)=\sum_{n_e,n_{ph}} \rho^{(n_e,n_{ph})}(t)$ give the probability that, during a certain time interval $t$, $n_{e}$ electrons have tunneled out of a given electron-phonon system and $n_{ph}$ phonons have been emitted\cite{qja}.
We define the generating function (GF)\cite{lenstra,dong}
$G(t,s_e,s_{ph})=\sum_{n_e,n_{ph}} s_e^{n_e} s_{ph}^{n_{ph}}\rho^{(n_e,n_{ph})}(t)$,
where $s_{e(ph)}$ are the electron (phonon) counting variables whose derivatives give us the correlations:
\begin{equation}\label{corr}
\frac{\partial^{p+q} trG(t,1,1)}{\partial s_e^p\partial s_{ph}^{q}}=\left\langle \prod_{i=1}^p\prod_{j=1}^{q}(n_e-i+1)(n_{ph}-j+1)\right\rangle.
\end{equation}
Thus, we are able to obtain the mean  
number $\langle n_\alpha\rangle$, the variance $\sigma_\alpha^2=\langle n_\alpha^2\rangle -\langle n_\alpha\rangle^2$ (which give the $\alpha=e,ph$ current and noise, respectively), or define the correlation between the electron and phonon counts, $\langle n_{e}n_{ph}\rangle$.

We integrate the equations of motion for the GF:
\begin{equation}\label{eomgf}
\dot G(t,s_e,s_{ph})=M(s_e,s_{ph})G(t,s_e,s_{ph}),
\end{equation}
that generalises 
the Master equation, $\dot\rho(t)=M(1,1)\rho(t)$, by introducing the counting variables in those terms corresponding to the tunneling of an electron to the collector lead and the emission of a phonon.

The longtime behaviour is 
extracted from the pole 
near $z=0$ in the Laplace transform of the GF, $\tilde G(z,s_e,s_{ph})=(z-M)^{-1}\rho(0)$.
From the Taylor expansion of the pole 
$z_0=\sum_{m,n>0}c_{mn}(s_e-1)^m(s_{ph}-1)^n$, we obtain 
$G(t,s_e,s_{ph})\sim {\textsl g}(s_e,s_{ph})e^{z_0t}$ and 
the central moments:
\begin{subequations}
\begin{equation}
\langle n_{e(ph)}\rangle=\frac{\partial {\textsl g}(1,1)}{\partial s_{e(ph)}}+c_{10(01)}t
\end{equation}
\begin{equation}
\sigma_{e(ph)}^2=\frac{\partial^2 {\textsl g}(1,1)}{\partial s_{e(ph)}^2}-\left(\frac{\partial {\textsl g}(1,1)}{\partial s_{e(ph)}}\right)^2+(c_{10(01)}+2c_{20(02)})t
\end{equation}
\end{subequations}
or the electron-phonon correlation (not discussed here), given by:
\begin{equation}
\langle n_{e}n_{ph}\rangle-\langle n_{e}\rangle\langle n_{ph}\rangle=\frac{\partial^2 {\textsl g}(1,1)}{\partial s_{e}\partial s_{ph}}+c_{11}t.
\end{equation}
Higher moments can be straightly forward obtained by this
formalism.
In the large time asymptotic limit, 
all the information is included in the coefficents $c_{mn}$. Then, the Fano factor is $F_{e(ph)}=1+2c_{20(02)}/c_{10(01)}$ so the sign of the second term in the right hand side defines the sub- ($F<1$) or super-($F>1$) Poissonian character of the noise. 

We describe our system by the Hamiltonian: 
$\hat H(t)=\sum_{i}\varepsilon_i\hat d_i^\dagger\hat d_i+\frac{\Omega}{2}\left(e^{-i\omega t}\hat d_2^\dagger\hat d_1+hc.\right)+\sum_Q\omega_Q\hat a_Q^\dagger\hat a_Q+\sum_Q\lambda_Q\left(\hat d_2^\dagger\hat d_1\hat a_Q+hc.\right)+\sum_{k\sigma}\varepsilon_{k\alpha}\hat c_{k\alpha}^\dagger\hat c_{k\alpha}+\sum_{k\alpha i}V_{\alpha}\left(c_{k\alpha}^\dagger\hat d_i + hc.\right),$ where $\hat a_Q$, $\hat c_{k\alpha}$ and $\hat d_i$ are annhilation operators of phonons and electrons in the leads and in the QD, respectively. 
Writing the density matrix as a vector, $\rho=(\rho_{00},\rho_{11},\rho_{12},\rho_{21},\rho_{22})^T$, the equation of motion of the GF (\ref{eomgf}) is described, in the Born-Markov approximation, 
by the matrix:
\begin{widetext}
\begin{equation}
\displaystyle
M(s_e,s_{ph})=\left(\begin{array}{ccccc}
-2\Gamma_{\rm L}-(\chi_1+\chi_2)\Gamma_{\rm R} & s_e\bar\chi_1\Gamma_{\rm R} & 0 & 0 & s_e\bar\chi_2\Gamma_{\rm R} \\
\Gamma_{\rm L}+s_e^{-1}\chi_1\Gamma_{\rm R} &  -\bar\chi_1\Gamma_{\rm R} & i\frac{\Omega}{2} & -i\frac{\Omega}{2} & s_{ph}\gamma  \\
0 & i\frac{\Omega}{2} & -\frac{(\bar\chi_1+\bar\chi_2)\Gamma_{\rm R}+\gamma}{2}+i\Delta_{\omega} & 0 & -i\frac{\Omega}{2} \\
0 & -i\frac{\Omega}{2} & 0 &-\frac{(\bar\chi_1+\bar\chi_2)\Gamma_{\rm R}+\gamma}{2}-i\Delta_{\omega} & i\frac{\Omega}{2} \\
\Gamma_{\rm L}+s_e^{-1}\chi_2\Gamma_{\rm R} & 0 & -i\frac{\Omega}{2} & i\frac{\Omega}{2} & -\gamma-\bar\chi_2\Gamma_{\rm R}\\
\end{array}  \right),
\end{equation}
\end{widetext}
where $\gamma=2\pi|\lambda_{\varepsilon_2-\varepsilon_1}|^2$ is the spontaneous phonon emission rate,
$\Gamma_{\alpha}=2\pi|V_\alpha|^2$ is the tunneling rate through the contact $\alpha$
, $\Omega$ is the Rabi frequency, which is proportional to the intensity of the ac field,
and $\chi_i=f(\varepsilon_i-\mu)=\left(1+e^{(\varepsilon_i-\mu)\beta}\right)^{-1}$ and $\bar{\chi}_i=1-\chi_i$ 
weight the tunneling of electrons between the right lead (with a chemical potential $\mu$) and the state $i$ in the QD. The Fermi energy of the left lead is considered infinite, so no electrons can tunnel from the QD to the left lead.
All the parameters in these equations, except the
sample-depending
coupling to the phonon bath,
can be externally manipulated. In the limit 
$\Gamma_{L(R)}\rightarrow 0$ we recover the pure 
RF case for the statistics of the emitted phonons\cite{lenstra},
\begin{equation}\label{frf}
F_{ph}(\Gamma_i=0)=1-\frac{2\Omega^2(3\gamma^2-4\Delta_\omega^2)}{(\gamma^2+2\Omega^2+4\Delta_\omega^2)^2},
\end{equation}
yielding the famous
sub-Poissonian noise result at resonance ($\Delta_\omega=0$).
In the following, we will restrict ourselves to the resonant case.

\begin{figure}[htb]
\begin{center}
\includegraphics[width=2.2in,clip]{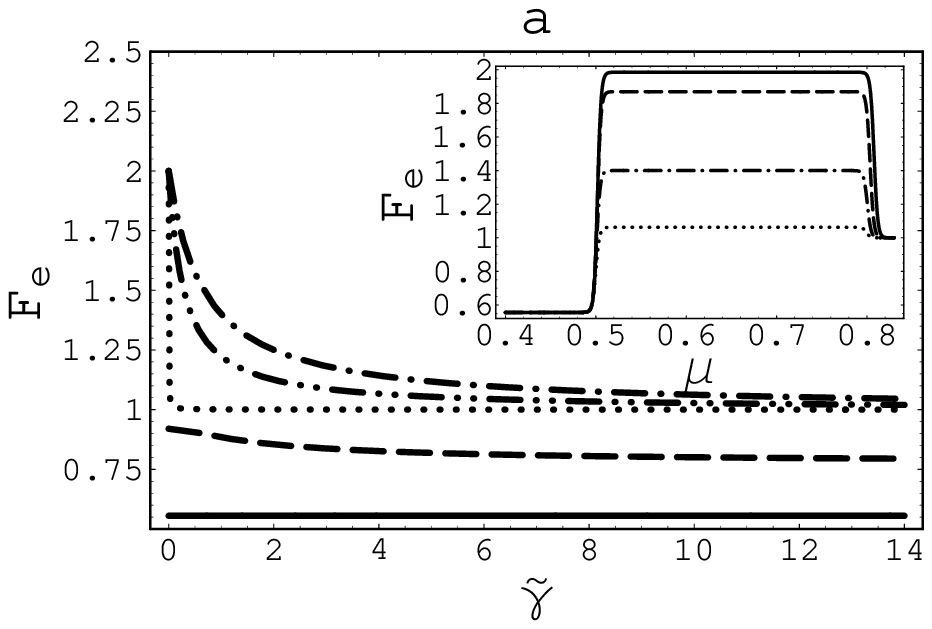}
\includegraphics[width=2.2in,clip]{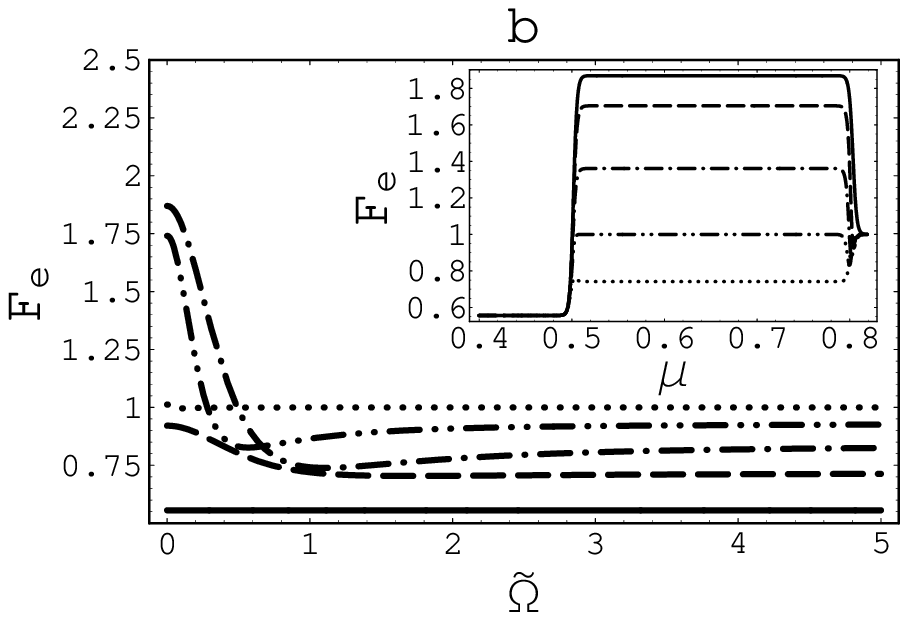}
\end{center}
\caption{\label{fanovsmuRg}\small Electron Fano factor as a function of (a) $\tilde\gamma=\gamma/\Gamma$ for $\tilde\Omega=\Omega/\Gamma=0$ ($\Gamma=\Gamma_{L/R}$) and (b) $\tilde\Omega$ for $\tilde\gamma=0.1$ for different chemical potentials: $\mu<\varepsilon_1$ (solid), $\mu=\varepsilon_1$ (dashed), $\varepsilon_1<\mu<\varepsilon_2$ (dash-dotted), $\mu=\varepsilon_2$ (dash-dot-dotted) and $\mu\gtrsim\varepsilon_2$ (dotted). Note that, in the non-driven case (a), the phonon bath does not affect the sub or super-Poissonian character of the electron noise which can be manipulated introducing the resonant driving field (b). Insets: (a) $F_e$ as a function of $\mu$ for $\Omega=0$ and different phonon emission rates: $\tilde\gamma=0.01$ (solid), $\tilde\gamma=0.1$ (dashed), $\tilde\gamma=1$ (dash-dotted) and $\tilde\gamma=10$ (dotted) and (b) for $\tilde\gamma=0.1$ and different driving intensities: $\tilde\Omega=0$ (solid), $\tilde\Omega=0.15$ (dashed), $\tilde\Omega=0.3$ (dash-dotted), $\tilde\Omega=\tilde\Omega_{P}=0.492$ (dash-dot-dotted), $\tilde\Omega=1$ (dotted). Parameters (holding for all figures, in meV): $\varepsilon_1=0.5$, $\varepsilon_2=0.8$, $\kappa_BT=\beta^{-1}=2\times10^{-3}$.
}
\end{figure}

{\em Electron noise--}
By tuning $\mu$, we 
control the tunneling of electrons from the two levels. Let us first consider the {\it non-driven} case ($\Omega=0$). 
If $\mu<\varepsilon_2$ (i.e., $\chi_2\approx0$), the Fano factor becomes:
\begin{widetext}
\begin{equation}\label{fanomu2}
F_e=1+\frac{2 \Gamma _L \Gamma _R \left(-2 \left(\gamma +\Gamma _R\right)^2+\Gamma _L \Gamma _R \chi
   _1^2+\left(2 \gamma  \Gamma _L+\left(\gamma +\Gamma _R\right) \left(2 \gamma +3 \Gamma
   _R\right)\right) \chi _1\right)}{\left(\left(\gamma +\Gamma _R\right) \left(2 \Gamma _L+\Gamma
   _R\right)-\Gamma _L \Gamma _R \chi _1\right)^2}.
\end{equation}
\end{widetext}
The {\it dynamical channel blockade} case is of special interest, as for  $\varepsilon_1<\mu<\varepsilon_2$ ($\chi_1\approx 1$) the occupation of the lower level blocks the transport through the upper level, leading to super-Poissonian noise\cite{superPoisson}:
\begin{equation}\label{fano1mu2}
F_e=1+\frac{2\Gamma_L\Gamma_R}{\Gamma_R\left(\gamma+\Gamma_R\right)+\Gamma_L\left(2\gamma+\Gamma_R\right)},
\end{equation}
which has a finite well-defined value in spite of the strong
current suppression.
In contrast, the large bias case, $\mu\ll\varepsilon_1$ ($\chi_1\approx\chi_2\approx0$), where the conduction through both channels is allowed, results in sub-Poissonian noise
\cite{djuricJAP}:
\begin{equation}\label{mu1}
F_e=1-\frac{4 \Gamma _L \Gamma _R}{\left(2 \Gamma _L+\Gamma _R\right)^2},
\end{equation}
independent of the coupling to the phonon bath.
Note that by comparing with the single resonant level case\cite{srl}, the inclusion of an additional level in the bias window increases the noise.
In the opposite limiting case, $\mu\gtrsim\varepsilon_2$ ($\chi_2\lesssim1$), where the transport is still possible due to the thermal
smearing
of the Fermi 
function, the noise tends to be Poissonian before the current is
completely suppressed for higher $\mu$ (cf. Fig. \ref{fanovsmuRg}).
The phonon bath mainly affects the current noise through the
non-driven device in the region
$\varepsilon_1\lesssim\mu\lesssim\varepsilon_2$ by
reducing
it as $\gamma$ increases
(`noise reduction by noise')
without modifying its super-Poissonian character, cf. Eq.
(\ref{fano1mu2}), and in a extremely sensible way in the regime
$\mu\gtrsim\varepsilon_2$, where the small rate for electrons
tunneling from the upper level to the right lead is not able to
compete with the phonon-induced relaxation that blocks the current
and suppresses the super-Poissonian noise (Fig. \ref{fanovsmuRg}).

The introduction of a finite ac field affects the character of the
electron noise only in the region where $\mu$ is between the
energies of the two levels (see Fig. \ref{fanovsmuRg}). The
possibility of pumping electrons from the lower to the upper level
by means of the absorption of one photon suppresses the dynamical
channel blockade and, 
consequently (at sufficiently large Rabi frequencies), the electronic noise turns out to be sub-Poissonian:
\begin{widetext}
\begin{equation}
F_e=1+
\frac{2\Gamma_L\Gamma_R\left(
\Gamma_L\left(2\gamma +\Gamma_R\right)(\gamma+\Gamma_R)^2+
\Gamma_R \left(\gamma+\Gamma_R\right)^3
-\Omega^2\left(2\left(\gamma^2+2\Omega^2\right)+4\Gamma_L\left(2\gamma +\Gamma_R\right)+
\Gamma_R\left(7\gamma+5\Gamma_R\right)\right)
\right)}
{\left(\Gamma_R\left((\gamma+\Gamma_R)^2+3\Omega^2\right)+\Gamma_L \left((2\gamma+\Gamma_R)(\gamma+\Gamma_R)+4\Omega^2\right)\right)^2}.
\end{equation}
\end{widetext}
For $\mu\ll\varepsilon_1$, the Fano factor does not depend on the
ac field and expression (\ref{mu1}) is recovered.

\begin{figure}[htb]
\begin{center}
\includegraphics[width=2.2in,clip]{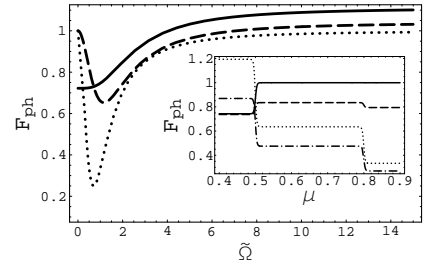}
\end{center}
\caption{\label{fanovsRabimugph}\small $F_{ph}$ as a function of $\tilde\Omega$ %
for $\tilde\gamma=1$ and different chemical potentials: $\mu<\varepsilon_1$ (solid), $\varepsilon_1<\mu<\varepsilon_2$ (dashed) and $\mu>\varepsilon_2$ (dotted). In the large bias case, the noise always increases with $\tilde\Omega$ and becomes super-Poissonian (in this concrete configuration) for $\tilde\Omega>4.046$. In the case $\mu>\varepsilon_2$,
there is a pronunced minimum at $\tilde\Omega=1/\sqrt{2}$ typical
of the RF. The case $\varepsilon_1<\mu<\varepsilon_2$ follows an
intermediate behaviour showing the RF-like minimum and the
super-Poissonian noise (typical of the large-bias case) for $\tilde\Omega>5\sqrt{2}$ (see text).
Inset: $F_{ph}$ as a function of $\mu$ for $\tilde\gamma=10$ and
different Rabi frequencies: $\tilde\Omega=0.1$ (solid),
$\tilde\Omega=2$ (dashed), $\tilde\Omega=6$ (dash-dotted),
$\tilde\Omega=10$ (dotted). For high $\tilde\Omega$, $\mu$ changes
the character of the noise. }

\end{figure}

{\em Phonon noise--} We now turn to 
the phonon statistics, cf. Fig. \ref{fanovsRabimugph}. In the {\it non-driven} case,
a 
phonon can be emitted only when an electron enters the upper level with 
the lower level
being already empty by an electron having tunneled out into the right contact. Therefore, the emission of phonons is anti-bunched and 
suppressed in the region where the transport of electrons is blocked by the occupation of the lower level (i.e., when $\mu>\varepsilon_1$):
\begin{equation}
F_{ph}=1-\frac{2\gamma\Gamma_L\Gamma_R\left(\gamma+2\Gamma_L+2\Gamma_R\right)\left(1-\chi_1\right)}
{(\Gamma_L(2\gamma+\Gamma_R(2-\chi_1))+\Gamma_R(\gamma+\Gamma_R))^2}
\end{equation}
However, the driving field gives an additional way of populating the higher level, increasing the noise and leading, for high enough field intensities, to the bunching of phonons (see Fig. \ref{fanovsRabimugph}).
A crossover from this {\it pure transport} infinite bias regime ($\mu\ll\varepsilon_1$), where the phonon noise increases with the Rabi frequency, to the strictly sub-Poissonian noise where the electron transport is suppressed (and the RF limit (\ref{frf}) is recovered) can be observed by increasing $\mu$ and successively blocking the transport through the lower (for $\mu>\varepsilon_1$) and upper level (for $\mu>\varepsilon_2$), cf. Fig. \ref{fanovsRabimugph}.
It can be easily seen that the noise suppression for this last case is largest 
when $\Omega=\gamma/\sqrt{2}$.
Thus, in the more interesting intermediate regime, $\varepsilon_1<\mu<\varepsilon_2$, 
as in the RF case, a minimum at low intensities is observed which is shifted and modified by the electronic transport, and also the super-Poissonian noise characteristic of the large bias regime for $\Omega>(4 \left(3\gamma+2\Gamma_R\right)\Gamma_L^2/\Gamma_R+\left(13\gamma+11\Gamma_R\right)\Gamma_L+3\Gamma_R\left(\gamma+\Gamma_R\right))^{1/2}$ (see Fig. \ref{fanovsRabimugph}).

\begin{figure}[htb]
\begin{center}
\includegraphics[width=2.1in,clip]{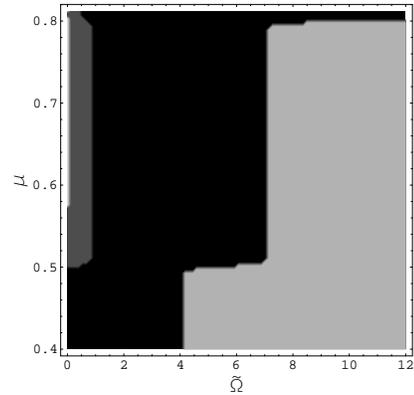}
\end{center}
\caption{\label{fanoareas}\small Colour plot showing the different regions where one can find $F_e,F_{ph}\ge1$ (white, appearing only for $\tilde\Omega=0$), $F_e<1, F_{ph}\ge1$ (light grey), $F_e\ge 1, F_{ph}<1$ (dark grey) and $F_e,F_{ph}<1$ (black) by tuning $\mu$ and $\tilde\Omega$, for $\tilde\gamma=1$.
}
\end{figure}

Joining both studies, it is possible to find different regions where the electron and phonon noise show all the sub and super-Poissonian combinations except the bunching of both electrons and phonons, since in the low intensities regime, where the electron noise tends to be super-Poissonian, the phonon noise is sub-Poissonian (strictly Poissonian for $\Omega=0$), see Fig. \ref{fanoareas}.


In conclusion, we propose a solid-state, transport version of resonance fluorescence (driven two-level quantum dot with phonon emission), where
the noise for both the transferred electronic current and the emitted phonons can be experimentally controlled
by tuning
the transport (chemical potential\cite{eth}) and optics (ac field intensity) parameters. We found various  
regimes showing different combinations of sub and super-Poissonian electron and phonon Fano factors.
Although not discussed here, our technique  can also be applied to driven electron-phonon systems with higher complexity such as double quantum dots \cite{BAP04}, and to obtain higher moments and correlations between electron and phonon distributions, 
which will be the scope of a future work.

We thank E. Sch\"oll and G. Kie\ss lich for discussions. Work supported by the M.E.C. of Spain
through Grant No. MAT2005-00644 and the UE network Grant No. MRTN-CT-2003-504574. R.S. was supported by CSIC-Programa I3P, cofinanced by Fondo Social Europeo.

\end{document}